\documentclass[conference,a4paper]{IEEEtran}
\usepackage[a4paper, left=44pt, right=44pt, bottom=72pt, top=54pt]{geometry}
  
\IEEEoverridecommandlockouts
\usepackage{cite}
\usepackage{float}
\usepackage{amsmath,amssymb,amsfonts}
\usepackage{graphicx}
\usepackage{balance}
\usepackage{amsthm}
\usepackage{mathtools}
\usepackage[short]{optidef}
\usepackage[normalem]{ulem}

\usepackage{xcolor}

\newcommand\norm[1]{\left\lVert#1\right\rVert}
\renewcommand{\t}{^\top}
\newcommand{\mmatrix}[1]{\begin{bmatrix}#1\end{bmatrix}}

\def\BibTeX{{\rm B\kern-.05em{\sc i\kern-.025em b}\kern-.08em
    T\kern-.1667em\lower.7ex\hbox{E}\kern-.125emX}}

\usepackage{tikz}

\newcommand\acceptedtext{%
  \footnotesize © 2023 IEEE.  Personal use of this material is permitted.  Permission from IEEE must be obtained for all other uses, in any current or future media, including reprinting/republishing this material for advertising or promotional purposes, creating new collective works, for resale or redistribution to servers or lists, or reuse of any copyrighted component of this work in other works.}

\newcommand\acceptednotice{%
\begin{tikzpicture}[remember picture,overlay]
\node[anchor=south,yshift=15pt] at (current page.south) {\fbox{\parbox{\dimexpr0.65\textwidth-\fboxsep-\fboxrule\relax}{\acceptedtext}}};
\end{tikzpicture}
}

\begin{document}

\definecolor{RoyalBlue}{cmyk}{1, 0.50, 0, 0}
\definecolor{ForestGreen}{cmyk}{0.41 0 0.41  0.45}

\title{Autonomous Path Following using Data-Driven Predictive Control}

\title{Autonomous Path Following using Data-Driven Predictive Control\\
\thanks{This work has been supported in part by the Croatian Science Foundation under the project UIP-2019-04-6487.}
}

\author{\IEEEauthorblockN{Josip Kir Hromatko}
\IEEEauthorblockA{\textit{University of Zagreb, Faculty of} \\
\textit{Electrical Engineering and Computing} \\
Zagreb, Croatia \\
josip.kir.hromatko@fer.hr}
\and
\IEEEauthorblockN{Marko \v{S}vec}
\IEEEauthorblockA{\textit{University of Zagreb, Faculty of} \\
\textit{Electrical Engineering and Computing} \\
Zagreb, Croatia \\
marko.svec@fer.hr}
\and
\IEEEauthorblockN{\v{S}andor Ile\v{s}}
\IEEEauthorblockA{\textit{University of Zagreb, Faculty of} \\
\textit{Electrical Engineering and Computing} \\
Zagreb, Croatia \\
sandor.iles@fer.hr}
}

\maketitle
\acceptednotice

\begin{abstract}
Predictive control based on an informative system trajectory, instead of a physics-based model, has received significant attention in recent years. This paper investigates the potential of using such data-driven control for vehicle dynamics control and autonomous path following. By considering the path following problem in the error space, the underlying system is approximately linear and existing results for data-driven predictive control can be applied. Also, scheduling based on longitudinal speed can be readily included. The proposed control algorithm was tested on two different lane change maneuvers in a high-fidelity simulation environment.
\end{abstract}

\begin{IEEEkeywords}
fundamental lemma, active steering, model-free predictive control
\end{IEEEkeywords}

\section{Introduction}

Modern vehicles have become prime examples of complex mechatronic systems with various sensors, onboard computers, and electronic systems. With the goal of increasing safety, comfort, fuel economy, and performance, these systems are implemented using increasingly advanced algorithms.

As computational hardware and software tools have advanced, optimization problems can now be solved more efficiently. This development has made it possible to apply computationally demanding control strategies such as model predictive control (MPC) in various applications \cite{schwenzer2021review} including fast mechatronic systems such as vehicle dynamics \cite{siampis2018, wang2020autonomous, guo2020real, fu2022nmpc}.

The main benefits of using MPC are the systematic handling of state or input constraints and the simplicity of designing multiple-input multiple-output (MIMO) controllers. MPC is based on iterative optimization: at each sampling instant, the system model is used to find the best possible inputs and outputs (predictions) over a finite horizon \cite{mayne2014model}. One of the challenges in MPC is obtaining an accurate system model. To alleviate this problem or remove the need for an explicit model altogether, a purely data-driven control method has received attention in recent years \cite{markovsky2008,hou2013}. Instead of using a classical (state-space) system model, this approach relies on an informative system trajectory, i.e., a sequence of inputs and outputs. Contrary to other data-driven methods such as system identification or artificial neural networks, this method combines identification and prediction into one optimization problem which is repeatedly solved online, as in standard MPC schemes. Recent work by several research groups demonstrated both theoretical guarantees \cite{allgower_affine,berberich2021} and practical insights \cite{DeePC,lygeros_quad}. 
While some of these studies have considered nonlinear systems, the majority of the research has focused on linear and time-invariant (LTI) systems. However, there are researchers who have extended these findings to explore the application of data-driven predictive control in the context of input-output linear parameter-varying (LPV) systems \cite{verhoek2021data} that can serve as a bridge between linear and nonlinear systems.

Although vehicle dynamics is a well-researched area and accurate physical models of varying complexity exist, changing environmental conditions or vehicle parameters might introduce additional errors in the prediction model used for MPC. Moreover, in many control applications, a simplified vehicle dynamics model is typically used, and linearization techniques might be employed. System identification can be performed to find the optimal parameters for the predetermined model structure.

The problem of predictive autonomous vehicle path following (sometimes also called active steering) is a widely researched topic with many possible applications such as urban navigation or autonomous racing (see for example \cite{Borrelli2005MPCBasedAT, yoon2009} and \cite{lee2013}). The main goal of the active steering system is to enhance the steering response, stability, and overall handling of the vehicle. It can be used for lane keeping and obstacle avoidance \cite{turri2013linear, bian2019advanced}. In the aforementioned papers, both implicit and explicit MPC was reported. Most of the present literature on lateral vehicle dynamics control relies on simplified planar nonlinear vehicle models, often represented as a bicycle model coupled with nonlinear tire models, or relies on a linearized model that depends on the vehicle speed. Some authors include longitudinal and lateral load transfer, while others neglect it for control design purposes. The simplifications made during modeling can influence the controller design.

This paper presents a data-driven predictive control algorithm for autonomous vehicle path following. It relies on measurements of the vehicle's longitudinal and lateral speed, yaw rate, steering wheel angle, and upcoming road curvature. In contrast with the commonly used vehicle dynamics control methods, no further knowledge about the vehicle or the environment is assumed. The control input is the steering wheel angle, while the longitudinal speed is controlled by an external module. Comparison with the commonly used linearized velocity-dependent bicycle model of the vehicle in error space is included and the results are compared in a high-fidelity simulation environment.

The paper is structured as follows: In Section \ref{ch:vehicle_model}, a linear bicycle model of the vehicle is described, while Section \ref{ch:DDPC} presents the proposed control algorithm. Section \ref{ch:sim_env} describes the simulation environment used for testing and Section \ref{ch:results} covers the presentation and discussion of the obtained results. Finally, Section \ref{ch:conclusion} provides some concluding remarks.

\section{Vehicle model}
\label{ch:vehicle_model}

For the path following problem, we consider a bicycle vehicle model described in \cite{rajamani}. This model assumes that two wheels per axle are lumped into one and that the vehicle is actuated by front wheel steering, as shown in Figure \ref{fig:bicycle}. The continuous-time lateral vehicle dynamics can then be represented as a linear parameter-varying state-space model: 
\begin{figure}[b]
    \centering
    \includegraphics[width=0.7\columnwidth]{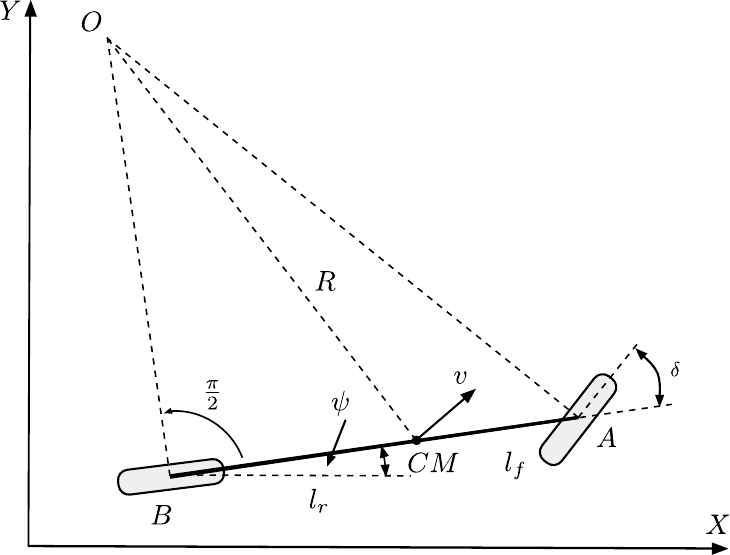}
    \caption{The bicycle model of a vehicle.}
    \label{fig:bicycle}
\end{figure}

\begin{equation}\label{eq:xp}
    \begin{aligned}
    \dot{x}_p &= A_p(v_x)x_p + B_p\delta \\
    y_p &= C_px_p
    \end{aligned}
\end{equation}
where $x_p=[v_y\ \dot{\psi}]\t$, $C_p$ is the identity matrix and the remaining matrices are:
\begin{equation}
\resizebox{.9\columnwidth}{!}{$
    A_p(v_x)=\mmatrix{-\frac{2C_f+2C_r}{mv_x} & -v_x-\frac{2C_fl_f-2C_rl_r}{mv_x}\\
                -\frac{2C_fl_f-2C_rl_r}{I_zv_x} & -\frac{2C_fl_f^2+2C_rl_r^2}{I_zv_x}},\ 
    B_p=\mmatrix{\frac{2C_f}{m} \\ \frac{2C_fl_f}{m}}
$}
\end{equation}
Here, $v_y$ denotes the lateral vehicle speed in the body frame, $\dot{\psi}$ is the vehicle's yaw rate and $\delta$ is the steering angle of the front wheels. Front and rear tire cornering stiffness is denoted by $C_f$ and $C_r$, $m$ is the vehicle mass, $I_z$ the moment of inertia about the vertical axis, $v_x$ the longitudinal vehicle speed in the body frame. The distances of the center of mass from the front and rear axles are denoted by $l_f$ and $l_r$.

If the aim is to follow the center of the lane, the system can be described in terms of the lateral position and orientation errors, $e_1$ and $e_2$, defined in \cite{rajamani} by their derivatives:
\begin{equation}
    \dot{e}_1 = v_xe_2 + v_y, \quad \dot{e}_2 = \dot{\psi}-\dot{\psi}_{des}
\end{equation}
where $\dot{\psi}_{des}$ is the desired yaw rate of the vehicle. The error state dynamics can be related to $x_p$ in \eqref{eq:xp} as:
\begin{equation}
    \mmatrix{\dot{e}_1\\\dot{e}_2} = \underbrace{\mmatrix{0&v_x\\0&0}}_{A_e(v_x)} \mmatrix{e_1\\e_2} + \mmatrix{1&0&\hphantom{-}0\\0&1&-1} \mmatrix{x_p\\\dot{\psi}_{des}}
\end{equation}
Finally, the resulting state-space model from $u=[\delta\ \dot{\psi}_{des}]\t$ to $e=[e_1\ e_2]\t$ becomes:
\begin{equation}\label{eq:error_ss}
\resizebox{.9\columnwidth}{!}{$
\begin{aligned}
    \mmatrix{\dot{v}_y\\\ddot{\psi}\\\dot{e}_1\\\dot{e}_2} &= \mmatrix{A_p(v_x)&0_{22}\\I_2&A_e(v_x)} \mmatrix{v_y\\\dot{\psi}\\e_1\\e_2} + \mmatrix{B_p&0_{21}\\0&0\\0&\llap{$-$}1}\mmatrix{\delta\\\dot{\psi}_{des}} \\
    \mmatrix{e_1\\e_2}&=\mmatrix{0&0&1&0\\0&0&0&1}\mmatrix{v_y\\\dot{\psi}\\e_1\\e_2}
\end{aligned}
$}
\end{equation}
and the problem of path following translates to driving the system outputs $e_1$ and $e_2$ to zero. Notation $0_{mn}$ denotes a zero matrix of size $m\times n$, while $I_m$ denotes an identity matrix of size $m$.

With $x=[v_y\ \dot{\psi}\ e_1\ e_2]\t$, $p=v_x$ and $y=e$, the error state system model takes the standard linear parameter-varying (LPV) form:
\begin{equation}\label{eq:LTIc}
\begin{aligned}
    \dot{x}&=A(p)x+Bu\\
    y&=Cx
\end{aligned}
\end{equation}
Although the data-driven approach has been applied to LPV systems \cite{verhoek2021data}, in this work it is assumed that the longitudinal speed is constant. This assumption is reasonable if there is an upper-level controller (or the driver) that keeps the speed approximately constant. The matrix $A$ then becomes fixed and the system is linear time-invariant (LTI). A discrete-time version of the system can be obtained by discretizing the matrices in \eqref{eq:LTIc} with the chosen sampling time $T_s$.

\textit{Remark:} To avoid numerical errors due to integrator drift and ensure tracking of the global lateral position $Y$, a different expression from \cite{rajamani} for the lateral position error is used:
\begin{equation}\label{eq:e1mod}
    e_1=\frac{Y-Y_{des}}{\cos{\psi}}
\end{equation}
which is assumed to be approximately equal to the standard definition. In addition, the 2D position in the global coordinate frame can be obtained from the body-fixed measurements using the following expressions:
\begin{equation}
\begin{aligned}
    \dot{X}&=v_x\cos(\psi) - v_y\sin(\psi)\\
    \dot{Y}&=v_x\sin(\psi) + v_y\cos(\psi)
\end{aligned}
\end{equation}

\section{Data-driven predictive path following}\label{ch:DDPC}
\subsection{The fundamental lemma}
The core assumption of the trajectory-based system modelling approach is based on the so-called \textit{fundamental lemma} described in \cite{willems2004}, which can be summarised as follows.

Given a sequence of $N$ consecutive samples of a signal, $x=\{x_0,\ x_1,\ \dots,\ x_{N-1}\}$, the corresponding Hankel matrix of order $L$ is defined as:
\begin{equation}
    H_L(x) \coloneqq \mmatrix{x_0&x_1&\dots&x_{N-L}\\
    x_1&x_2&\dots&x_{N-L+1}\\
    \vdots&\vdots&\ddots&\vdots\\
    x_{L-1}&x_L&\dots&x_{N-1}}
\end{equation}
Additionally, an input sequence $u$, with $u_k\in\mathbb{R}^m$, is said to be persistently exciting of order $L$ if $\text{rank}(H_L(u))=mL$.

Assume that $u_d$ and $y_d$ represent the samples of a trajectory of a controllable discrete-time LTI system $G$ and $u_d$ is persistently exciting of order $L+n$, where $n$ is the state dimension of $G$. Then, $\bar{u}$ and $\bar{y}$ are a trajectory of $G$ if and only if there exists a vector $\alpha\in\mathbb{R}^{N-L+1}$ such that:
\begin{equation}
    \mmatrix{H_L(u_d)\\H_L(y_d)}\alpha=\mmatrix{\bar{u}\\\bar{y}}
\end{equation}
The results also hold if the assumed system dimension $n$ is replaced by an upper bound, in case of unknown true system dimensions.

\subsection{The receding horizon approach}
\subsubsection{Data-driven predictive controller}
Standard model predictive control schemes rely on explicit discrete-time system models. However, the fundamental lemma allows all possible system trajectories to be described directly from the collected data. This condition can then be used in a receding horizon scheme instead of the system dynamics constraint.

The data-driven predictive controller thus aims to repeatedly solve the following constrained optimization problem:
\begin{mini!}
    {u,y,\alpha,\sigma}{\lambda_\alpha\norm{\alpha}_2^2 + \lambda_\sigma\norm{\sigma}_2^2  + \sum_{k=0}^{L}\ell(u_k,y_k)\label{eq:opt_cost}}{\label{eq:opt}}{}
    \addConstraint{\mmatrix{u_{[-n,L-1]}\\y_{[-n,L-1]}}}{=\mmatrix{H_{L+n}(u_d)\\H_{L+n}(y_d)\label{eq:opt_model}}\alpha}{}
    \addConstraint{\mmatrix{u_{[-n,-1]}\\y_{[-n,-1]}}}{=\mmatrix{u_{ini,k}\\y_{ini,k}} + \mmatrix{0\\\sigma}\label{eq:opt_ini}}{}
    \addConstraint{u_k\in\mathbb{U},}{\ \  y_k\in\mathbb{Y},\quad \label{eq:opt_lim}}{k=0,\ldots,L}
\end{mini!}
with $L$ being the prediction horizon and $n$ the assumed system dimension. Notation $x_{[a,b]}$ represents a stacked window, $x_{[a,b]}\coloneqq [x_a\t\ \ldots\ x_b\t]$. Constraint \eqref{eq:opt_model} now represents the system "model", while constraints \eqref{eq:opt_lim} include the control input and system output limitations. Constraint \eqref{eq:opt_ini} sets the initial condition and closes the feedback loop: instead of the standard constraint $x_0=x(t)$, where $x(t)$ is the current state measurement/estimate, the initial state is set implicitly through $n$ past input-output pairs ($u_{ini}$ and $y_{ini})$, removing the need for explicit state estimation.

The cost function \eqref{eq:opt_cost} describes the user-defined control objectives. It is common to use a quadratic function for the running cost $\ell$:
\begin{equation}\label{eq:running_cost}
    \ell(u,y) = y\t Qy + u\t Ru
\end{equation}
which penalizes both the state deviations and the control effort with the positive definite weight matrices $Q$ and $R$. Generally, a different cost and constraints could be defined for the last $n$ elements in the optimization sequence, corresponding to terminal ingredients in classic MPC schemes. If these are carefully chosen, certain closed-loop guarantees can be obtained. However, in this work it is assumed that practical stability can be achieved by using a sufficiently long prediction horizon, as shown in \cite{bongard2023}. The slack variable $\sigma$ is introduced to relax the initial condition constraint on the outputs in the presence of noisy measurements or equivalent unmeasured errors, but its weighting term $\lambda_\sigma$ should be chosen relatively large to ensure an accurate approximation to the initial conditions. In addition, the regularization term on $\alpha$ improves robustness to imperfect data and numerical stability, as discussed in \cite{coulson2019}.

With the quadratic cost function and linear constraints, \eqref{eq:opt} becomes a standard quadratic program (QP) which can be solved efficiently with existing QP solvers. Compared to the classic MPC, the size of the optimizer is increased due to the vector $\alpha$. However, with an appropriate choice of $N$ and $L$, similar computation times can be achieved.

\subsubsection{Model-based predictive controller}
For comparison, the standard linear time-invariant MPC controller is used:
\begin{mini!}
    {u}{\ell_T(y_{L+1})  + \sum_{k=0}^{L}\ell(u_k,y_k) \label{eq:opt_cost_LTI}}{\label{eq:opt_LTI}}{}
    \addConstraint{
    x_{k+1} = A_d x_k + B_d u_k}
    \addConstraint{
    y_k = Cx_k}
    \addConstraint{
    x_0 = x(t)}
    \addConstraint{ u_k \in\mathbb{U},\ y_k\in\mathbb{Y},}{}{\quad k=0,\ldots,L}
\end{mini!}
The matrices $A_d$ and $B_d$ in the state evolution equation are the discrete-time versions of the system matrices obtained by discretizing the continuous-time system given in \eqref{eq:LTIc}. The discretization is performed with a chosen sampling time $T_s$ assuming a constant vehicle speed $v_x$ over the prediction horizon, while $x(t)$ represents the initial condition (i.e., the current state estimate). The control input constraint limits the maximum allowed steering angle and sets the desired yaw rate (defined by the driver, i.e., the maneuver ). The weight for the running cost was set equal to \eqref{eq:running_cost} and a quadratic term, $y^\top Qy$, was used as the terminal cost.

\subsubsection{Identification-based predictive controller}
Finally, to allow a fair comparison (limited knowledge of the system model), a black-box state-space model was estimated using the same trajectory as for the data-driven controller. Assuming that the system order is known, the matrices $A_{id}$ and $B_{id}$ were obtained by a least squares fit, i.e., by solving the following optimization problem:
\begin{mini}
    {A_{id},B_{id}}{\sum_{k=0}^{N-2}\Vert x_{k+1} - (A_{id}x_k + B_{id}u_k)
    \Vert^2_2}{\label{eq:sysidAB}}{}
\end{mini}
where $x_k$ and $u_k$ denote the samples of a recorded trajectory of length $N$. In the case where only output measurements are available, one can introduce surrogate states by stacking consecutive outputs. The output matrix $C_{id}$ can be found by solving a similar optimization problem including the outputs $y_k$:
\begin{mini}
    {C_{id}}{\sum_{k=0}^{N-2}\Vert y_k - C_{id}x_k \Vert^2_2}{\label{eq:sysidC}}{}
\end{mini}

The problem \eqref{eq:sysidAB} can be solved analytically with a Moore-Penrose pseudoinverse denoted by $(\cdot)^\dagger$:
\begin{equation}
    \left[A_{id} \ B_{id} \right] = UV\t(VV\t)^{\dagger},
\end{equation}
with
\begin{equation*}
\begin{split}
     U &= \left[ x_1, \ x_2, \ \dots,\ x_{N-1}\right] \\
     V &= \left[ [x_0, u_0]\t, \ [x_1, u_1]\t,\ \dots,\ [x_{N-2}, u_{N-2}]\t \right]
\end{split}
\end{equation*}
Similarly, \eqref{eq:sysidC} can be solved as:
\begin{equation}
    C_{id} = YW\t (W W\t)^{\dagger},
\end{equation}
where
\begin{equation}
\begin{split}
    Y &= \left[y_0, \ y_1,\ \dots,\ y_{N-1} \right] \\ 
    W &= \left[x_0, \ x_1,\ \dots,\ x_{N-1} \right]
\end{split}
\end{equation}
Finally, the identified matrices can be used instead of $A_d$, $B_d$, and $C$ in a classic MPC scheme \eqref{eq:opt_LTI}.

\subsection{Reference generation}
The error states are affected by the desired yaw rate of the vehicle, as described in \eqref{eq:error_ss}. The desired yaw rate is calculated as $\dot{\psi}_{des}=v_x\kappa$, where $\kappa$ denotes the road curvature. Given a desired path defined by a sequence of global coordinates $X=\{X_0,\ X_1,\ \ldots\ ,\ X_D\}$ and $Y=\{Y_0,\ Y_1,\ \ldots\ ,\ Y_D\}$, the road curvature at each point can be calculated as follows:
\begin{equation}
    \kappa_i = \frac{\Delta X_i\cdot\Delta^2 Y_i}{\left(\Delta X_i^2 + \Delta Y_i^2\right)^{3/2}}
\end{equation}
where $\Delta^m X$ represents the $m$-th order approximate derivative of $X$ and $\Delta X_i=(X_{i+1}-X_i)/T_s$. The desired maneuver is thus defined by the sequence $\kappa=\{\kappa_0,\ \kappa_1,\ \ldots\ ,\ \kappa_D\}$. Additionally, the alternative error definition given in \eqref{eq:e1mod} requires the global lateral position reference, $Y_{des}$.

\section{Simulation environment}
\label{ch:sim_env}
The proposed control algorithm was tested using Simulink and IPG CarMaker, which implements a high-fidelity nonlinear vehicle model. The basic parameters of the selected generic vehicle are listed in Table \ref{tab:params}. The control algorithm required measurements of longitudinal and lateral vehicle speed (in the body-fixed frame), the vehicle's yaw rate, and the steering wheel angle.

\begin{table}
    \centering
    \caption{Vehicle parameters}
    \resizebox{\columnwidth}{!}{
    \begin{tabular}{c c c c }
         \hline
         Parameter & Value & Unit & Description\\
         \hline
          $m$ &   1600  & kg & mass of the vehicle\\
          $l_f,l_r$ &  1.311  & m & front/rear axle to CoG distance\\
          $I_z$ & 2394 & kg $\cdot$ m\textsuperscript{2} & moment of inertia around the vertical axis\\
          $C_f,C_r$ & 72705 & N/rad & front/rear tire lateral stiffness (approximated)\\
          $i_{sw}$ & 13 & - & steering ratio\\
          \hline
    \end{tabular}}
    \label{tab:params}
\end{table}

\subsection{Data collection}
\begin{figure}
    \centering
    \includegraphics[width=\columnwidth]{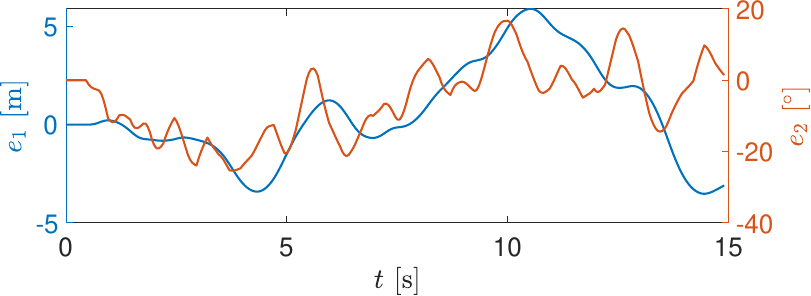}
    \caption{Data collected at 80 km/h.}
    \label{fig:PEtest}
\end{figure}

To create an informative dataset, random steering wheel inputs sampled uniformly in the range of $\pm 90^\circ$ at a sampling time of 0.25 seconds were added to CarMaker's driver commands during a straight drive at the desired speed for 15 seconds. The dataset size was chosen larger than needed to ensure there is enough information, but only the first 5 seconds were used in the final version of the controller. The lateral driver, which aims to follow the lane centerline, was used as a stabilizing controller to avoid increasing the error states too much and reducing the informativity of the dataset.  Additionally, the artificial control input (road curvature) was varied between $\pm 0.02\ \text{m}^{-1}$ to gain information about its effect on the error states. An example of measured error states for the speed of 80 km/h is shown in Figure \ref{fig:PEtest}. Persistently exciting datasets were collected at the speeds of 60, 80, and 100 km/h.

\subsection{Closed-loop testing}
The algorithm was tested by performing two driving maneuvers on dry asphalt at the speeds of 60, 80, and 100 km/h. Longitudinal speed control was done by CarMaker's longitudinal driver, while steering inputs were set by the predictive controller.

\subsubsection{Maneuver 1}
The first chosen maneuver was a double lane change described in \cite{Borrelli2005MPCBasedAT}, where the following equations describe the desired orientation and lateral position:
\begin{equation}
\resizebox{.88\columnwidth}{!}{$
\begin{aligned}
    \psi_{des}&=\frac{D_{y_1}}{2}(\tanh(Z_1)+1) - \frac{D_{y_2}}{2}(\tanh(Z_2)+1)\\
    Y_{des}&=\arctan\left(\Phi\right)\\
    Z_1&=\frac{S}{d_{x_1}}(X-X_{s_1})-\frac{S}{2}\\
    Z_2&=\frac{S}{d_{x_2}}(X-X_{s_2})-\frac{S}{2} \\
    \Phi&= \frac{1.2 D_{y_1}}{D_{x_1}}\left(\frac{1}{\cosh(Z_1)}\right)^2-\frac{1.2 D_{y_2}}{D_{x_2}}\left(\frac{1}{\cosh(Z_2)}\right)^2
\end{aligned}
$}
\end{equation}
with the parameters $S=2.4$, $D_{x_1}=25$, $D_{x_2}=21.95$, $D_{y_1}=4.05$, $D_{y_2}=5.7$, $X_{s_1}=27.19$ and $X_{s_2}=56.46$. 

\subsubsection{Maneuver 2}
The second maneuver is referred to as a multiple lane change and defined as:
\begin{equation}
\begin{aligned}
    Y_{des} = \frac{Y_{max}}{2} ( &\tanh(\rho X) - \tanh(\rho(X - dX_1)) \\
    - &\tanh(\rho (X - dX_1 - dX_2))\\
    + &\tanh(\rho (X - dX_1 - dX_2 - dX_3)) ) \\
    \psi_{des} = \arctan & (\dot{Y}_{des} / \dot{X})
\end{aligned}
\end{equation}
with the distance parameters $Y_{max} = 3$m, $dX_1 = 200$m, $dX_2 = 75$m, $dX_3 = 75$m and the curvature factor $\rho = 0.1\ \text{m}^{-1}$. The aim of this maneuver was to simulate overtaking in the first section and fast obstacle avoidance in the second.

\subsection{Controller parameters}
Table \ref{tab:ddpc_params} lists the parameters of the used data-driven predictive controller (DDPC). The weight matrices $Q$ and $R$ in \eqref{eq:running_cost} were chosen as diagonal, with zero weight on the desired yaw rate since it is fixed during the horizon. The same matrices and prediction horizon was used for the classic MPC and the identification-based controller. For the experiments, the steering wheel angle was limited to $\pm 70^\circ$ and the error states were unconstrained. The control algorithm was implemented using YALMIP \cite{lofberg2004yalmip} for the problem formulation and MOSEK \cite{aps2017mosek} for solving the resulting quadratic program.
\begin{table}
    \centering
    \caption{Controller parameters}
    \resizebox{\columnwidth}{!}{
    \begin{tabular}{c c c }
        \hline
        Parameter & Value & Description\\
        \hline
        $N$ & 100 & number of samples in the collected dataset\\
        $L$ & 12 & prediction horizon length\\
        $n$ & 4 & number of past measurements used \\
        $Q_{e_1}$ & 50 & lateral position tracking error weight\\
        $Q_{e_2}$ & 150 & orientation tracking error weight\\
        $R_{\delta_{SW}}$ & 0.5 & steering wheel angle weight\\
        $\lambda_\alpha$ & 0.25 & regularization term on $\alpha$\\
        $\lambda_\sigma$ & $1\cdot10^{6}$ & regularization term on the slack variable\\
        $T_s$ & $0.05$ s & control loop sampling time\\
        \hline
    \end{tabular}}
    \label{tab:ddpc_params}
\end{table}

\section{Results}
\label{ch:results}

\begin{figure}
    \centering
    \includegraphics[width=\columnwidth]{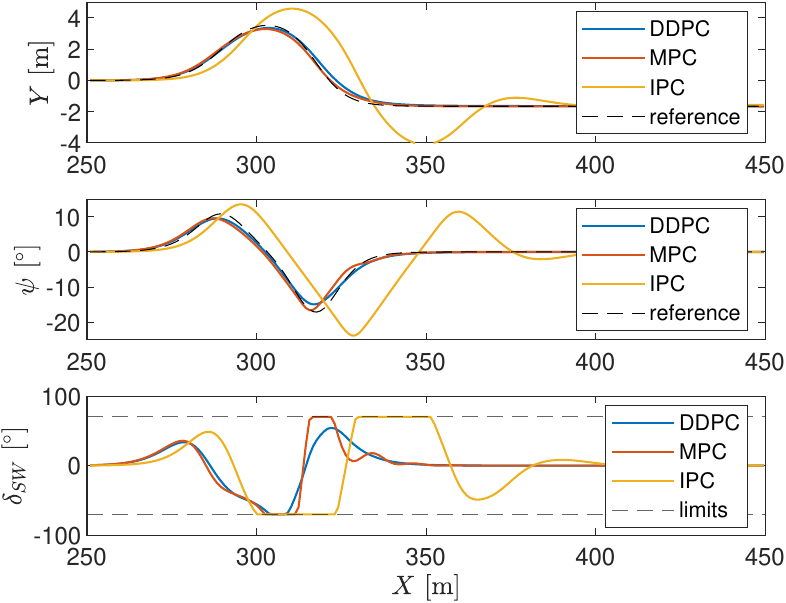}
    \caption{Comparison between the data-driven (DDPC), model-based (MPC) and identification-based (IPC) predictive controllers using the double lane change maneuver at 80 km/h.}
    \label{fig:Comparison}
\end{figure}

Figure \ref{fig:Comparison} shows the comparison between the proposed data-driven controller and classic model-based and identification-based controllers. The controllers were tested with the first test maneuver (double lane change) at a speed of 80 km/h. The results indicate that the nominal MPC and data driven MPC perform similarly, with a slightly larger solve time for the data-driven controller (not shown in the figure). On the other hand, the identification-based controller shows a larger deviation from the reference trajectory, although this result is not general and can vary depending on the identification method and tuning weights used.

\begin{figure}
    \centering
    \includegraphics[width=\columnwidth]{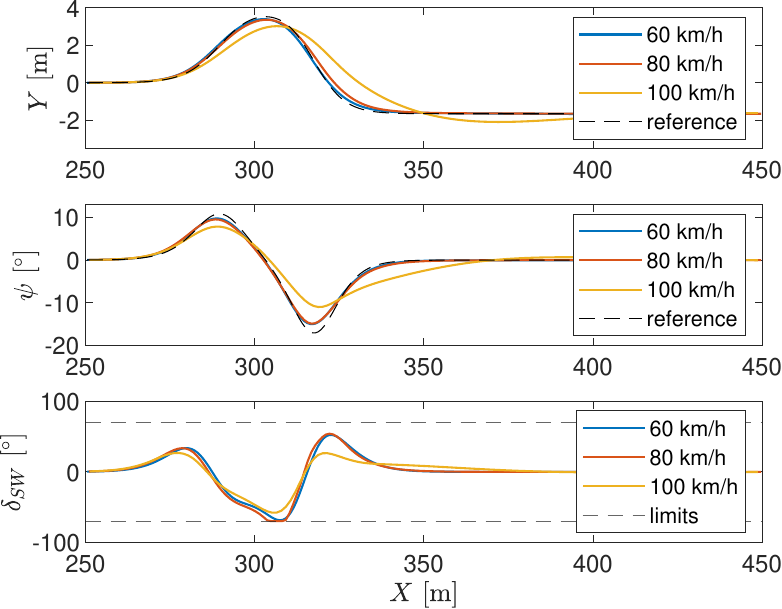}
    \caption{Path following during a double lane change using DDPC.}
    \label{fig:DLCcombined_ddpc}
\end{figure}

Figure \ref{fig:DLCcombined_ddpc} shows the results obtained with the data-driven controller when performing the double lane change at different speeds. The vehicle manages to follow the desired path relatively well at speeds of 60 and 80 km/h while respecting the imposed steering angle constraints. Performing the maneuver at 100 km/h appears to be more challenging, resulting in a slightly worse tracking performance. Additionally, the longitudinal speed was reduced by a maximum of 5 km/h in all three cases (not shown in the figures). Finally, it is important to note that the steering angle constraint was set to a lower value compared to the physical constraint.

\begin{figure}
    \centering
    \includegraphics[width=\columnwidth]{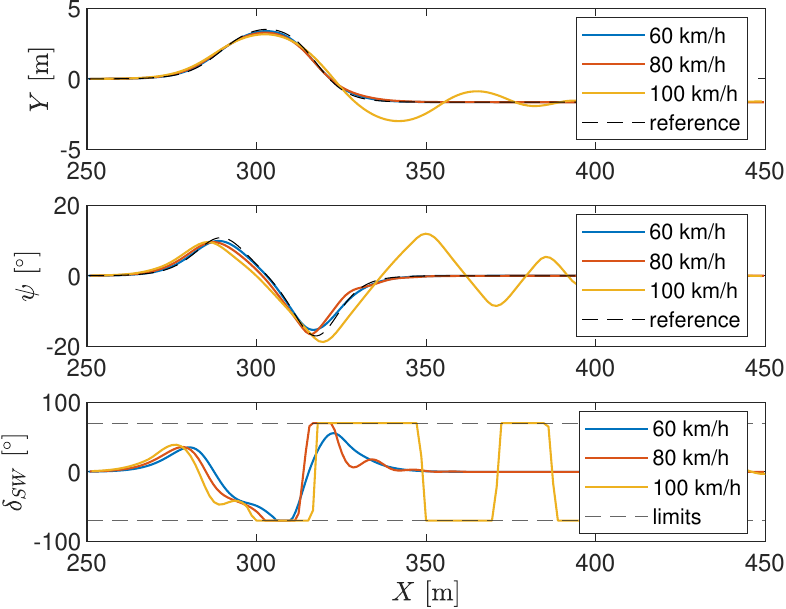}
    \caption{Path following during a double lane change using the nominal MPC.}
    \label{fig:DLCcombined_ltv}
\end{figure}

To make a comparison with the nominal MPC based on the linearized model and parameters obtained from CarMaker, the same test was repeated, and the results are shown in Figure \ref{fig:DLCcombined_ltv}. Since CarMaker uses a high-fidelity model that includes nonlinear effects that were neglected in the simplified control-oriented model, and since the nominal parameters were used, the performance deteriorated. However, changing the estimated tire cornering stiffness improved the tracking performance of the standard MPC. Additional tests for comparison to the linear MPC based on the linear model with identified parameters could be conducted. 

\begin{figure}
    \centering
    \includegraphics[width=\columnwidth]{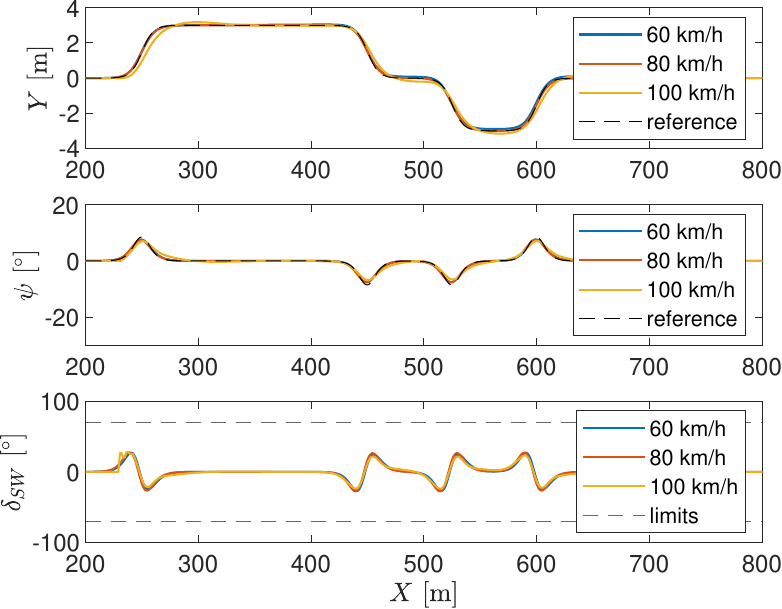}
    \caption{Path following during multiple lane changes using DDPC.}
    \label{fig:MLCcombined}
\end{figure}

The results for the second test maneuver and the data-driven controller are shown in Figure \ref{fig:MLCcombined}. In this case, the vehicle follows the desired path well at all three test speeds. Due to a smoother reference, the steering angle does not reach the imposed limits.

Mean and maximum solver times for all test cases were around 4 ms and 8 ms, respectively, using an Intel i7-7700, 3.6 GHz, 4 core, 16 GB memory machine. This suggests that the proposed algorithm should be able to run in real-time (under 50 ms) on a standard experimental setup.

On the other hand, it was observed that even a small change in certain parameters ($N$, $L$, $n$) could significantly deteriorate the tracking performance or lead to closed-loop instability. This could be caused by the implicit identification step inherent to the data-driven scheme, where longer prediction horizons might produce inaccurate predictions and poor control performance. To avoid this issue, one could try to reformulate \eqref{eq:opt} as a bi-level optimization problem, as proposed in \cite{lian2023adaptive}.

\section{Conclusion}
\label{ch:conclusion}
This paper presents a purely data-driven approach to autonomous vehicle path following using receding horizon control. Assuming no knowledge of the controlled vehicle and requiring only an informative sequence of commonly measured quantities, the data-driven predictive control was able to track a double lane change maneuver at different longitudinal speeds. The steering angle constraint was chosen to be less than physically possible to make the driving scenario more challenging. An additional experiment was conducted with multiple filtered step changes in the lateral position of the vehicle. In addition to demonstrating the applicability of the trajectory-based scheme to vehicle dynamics, several practical observations were also made. In particular, significant sensitivity to the tuning parameters was noticed. In future work, it would be interesting to perform a more detailed analysis of the effects of the parameters on the control performance and to investigate the possibility of a bi-level problem reformulation. Another aspect that should be considered is the investigation of LPV data-driven MPC techniques that would directly capture the dependence of the linear model on longitudinal speed. Finally, testing the proposed approach in an experimental setup would demonstrate its potential for practical applications. Since the presented approach does not require a model of the system, it can be easily extended to other vehicle dynamics control applications.
\balance

\bibliographystyle{IEEEtran}
\bibliography{bibliography}

\end{document}